\documentclass[10pt,aps,twocolumn,floats,floatfix,amssymb,prd,superscriptaddress,nofootinbib]{revtex4-2}

\usepackage{graphicx}
\usepackage{amsmath,amssymb}
\usepackage{amsfonts}
\usepackage{xspace}
\usepackage[usenames]{color}
\usepackage[dvipsnames]{xcolor}
\usepackage{dcolumn}
\usepackage{bm}
\usepackage{mathrsfs}
\usepackage[colorlinks=true]{hyperref}
\hypersetup{colorlinks=true,
            citecolor=NavyBlue,
            linkcolor=NavyBlue,
            urlcolor=NavyBlue}
\usepackage[all]{hypcap}
\usepackage[utf8]{inputenc}
\usepackage{multirow}
\usepackage{etoolbox}
\usepackage{tikz}
\usepackage{orcidlink}
\usepackage{microtype}
\usepackage[normalem]{ulem}

\usepackage[capitalise]{cleveref}

\newcommand{\GB}{\mathcal{G}}
\newcommand{\dd}{\textrm{d}}
\newcommand{\ii}{\textrm{i}}
\newcommand{\dV}{{\rm d}^{4}x \, \sqrt{-g}}

\crefformat{equation}{Eq.~(#2#1#3)}
\crefformat{table}{Tab.~(#2#1#3)}
\crefformat{figure}{Fig.~(#2#1#3)}
\crefformat{appendix}{App.~(#2#1#3)}
\crefformat{section}{Sec.~(#2#1#3)}
\crefformat{subsection}{Subsec.~(#2#1#3)}

\begin{document}

\title{Breaking black-hole uniqueness at supermassive scales}

\author{Astrid Eichhorn\,
\orcidlink{0000-0003-4458-1495}}
\email{eichhorn@sdu.dk}
\email{eichhorn@thphys.uni-heidelberg.de}
\affiliation{CP3-Origins, University of Southern Denmark, Campusvej 55, DK-5230 Odense M, Denmark}
\affiliation{Institut f\"ur Theoretische Physik, Universit\"at Heidelberg, Philosophenweg 16, 69120 Heidelberg, Germany}

\author{Pedro G. S. Fernandes\,\orcidlink{0000-0002-8176-7208}}
\email{pgsfernandes@sdu.dk}
\email{fernandes@thphys.uni-heidelberg.de}
\affiliation{CP3-Origins, University of Southern Denmark, Campusvej 55, DK-5230 Odense M, Denmark}
\affiliation{Institut f\"ur Theoretische Physik, Universit\"at Heidelberg, Philosophenweg 16, 69120 Heidelberg, Germany}

\author{Aaron Held\,\orcidlink{0000-0003-2701-9361}}
\email{aaron.held@phys.ens.fr}
\affiliation{
Institut de Physique Théorique Philippe Meyer, Laboratoire de Physique de l’\'Ecole normale sup\'erieure (ENS), Universit\'e PSL, CNRS, Sorbonne Universit\'e, Universit\'e Paris Cité, F-75005 Paris, France
}

\author{Hector O. Silva\,\orcidlink{0000-0002-0066-9471}}
\email{hector.silva@aei.mpg.de}
\affiliation{Max Planck Institute for Gravitational Physics (Albert Einstein Institute), D-14476 Potsdam, Germany}

\begin{abstract}
In general relativity, all vacuum black holes are described by the Kerr solution. Beyond general relativity, there is a prevailing expectation that deviations from the Kerr solution increase with the horizon curvature. We challenge this expectation by showing that, in a scalar-Gauss-Bonnet theory, black holes scalarize in a finite, adjustable window of black-hole masses, bounded from above and below. In this theory, there is an interplay between curvature scales and compactness, which we expect to protect neutron stars and other less compact objects from scalarization. In addition, this theory is the first to avoid the catastrophic instability of early-universe cosmology that affects previous scalarization models.
In this theory, black-hole uniqueness can be broken at supermassive black-hole scales, while stellar-mass black holes remain well-described by the Kerr solution. To probe this scenario, observations targeting supermassive black holes are necessary.
\end{abstract}

\maketitle

Our current best description of gravity, general relativity (GR), predicts that all vacuum black holes throughout the universe, spanning a mass range of at least ten orders of magnitude,
are
described by the Kerr metric~\cite{Kerr:1963ud}. This prediction of black-hole uniqueness is
referred to as the ``Kerr hypothesis", and is supported
by
uniqueness~\cite{PhysRevLett.26.331,PhysRevLett.34.905} and no-hair~\cite{Herdeiro:2015waa} theorems. Simultaneously,
stringent constraints on modifications of GR arise, because the predictions of GR agree to high accuracy
with observations in the weak-field regime at solar-system scales~\cite{Will:2014kxa}. This motivates
us to explore
modifications of
GR that give rise to
distinct
gravitational phenomena exclusively in the strong-field regime.
Within the strong-field regime, it is generically expected that the local spacetime curvature must exceed some threshold in order for deviations from GR to appear.
In this letter we challenge this expectation within the framework of black-hole \emph{scalarization}~\cite{Damour:1993hw,Damour:1996ke,Doneva:2017bvd,Silva:2017uqg}, see Ref.~\cite{Doneva:2022ewd} for a review.

\begin{figure}[t]
\includegraphics[width=\columnwidth]{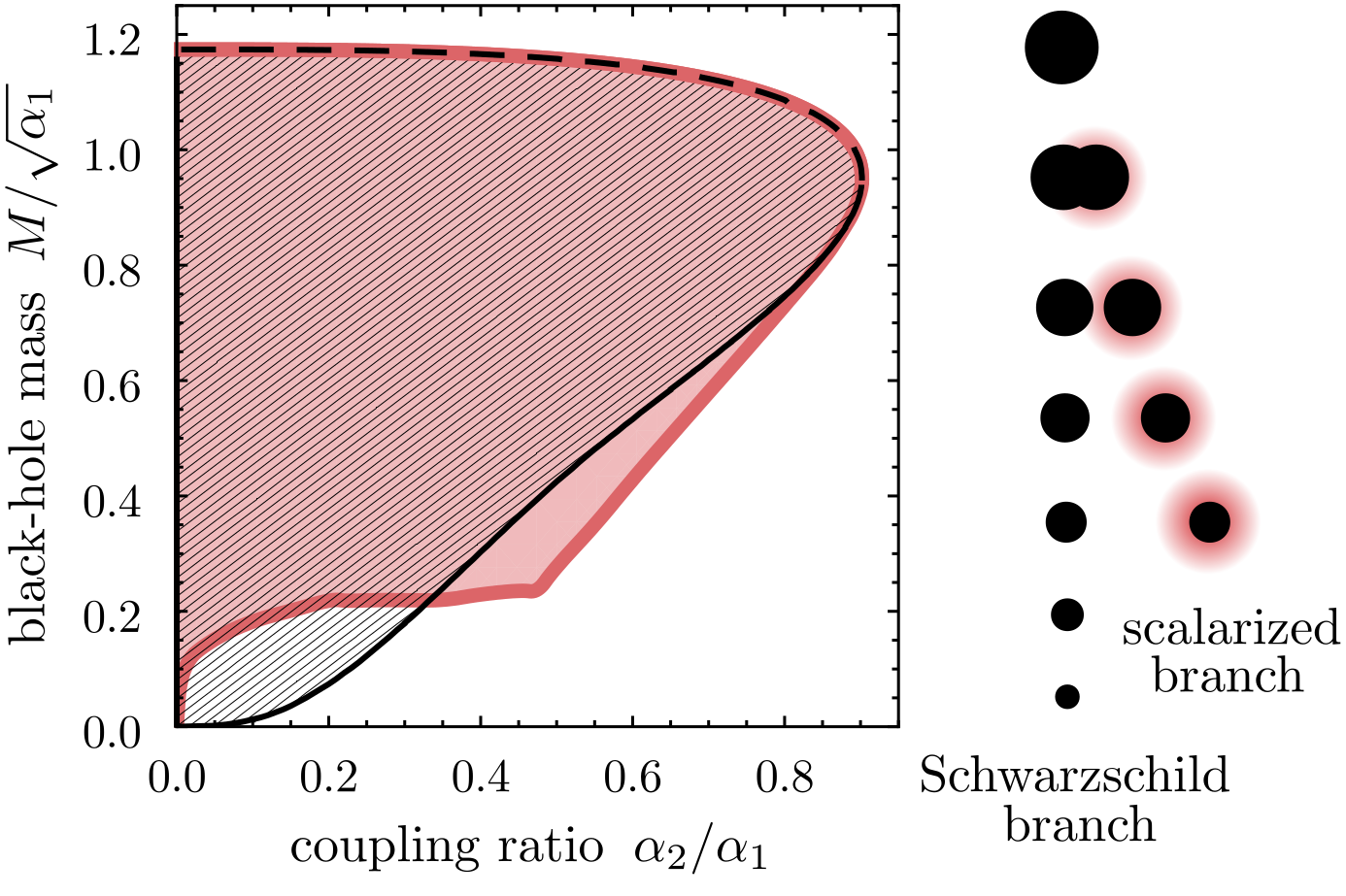}
\caption{Instability region of Schwarzschild spacetime (hatched region) and the
region in which we find scalarized black holes (red shaded region) for the
theory in Eq.~\eqref{eq:actionST}. At the upper end of the mass window, the
scalarized solution branches off the Schwarzschild branch continuously (as
indicated by the dashed curve). At the lower end, there is no continuous
transition between the two black-hole configurations.}
\label{fig:mplot}
\end{figure}

Black-hole scalarization happens in theories in which a scalar field couples to
the Gauss-Bonnet invariant, $\mathcal{G}$, a particular combination of the
Riemann curvature tensor and its contractions, inducing a tachyonic
instability around Kerr black holes under certain conditions, depending on the
black hole's mass~\cite{Doneva:2017bvd,Silva:2017uqg}, and its
spin~\cite{Dima:2020yac,Herdeiro:2020wei,Berti:2020kgk}.
Such an instability signals the emergence of a new branch of \emph{scalarized}
solutions, violating black-hole uniqueness.
In conventional scenarios, the instability affects all Kerr black holes below a
critical mass scale, and also neutron stars.
Observations of black holes in the range of a few solar masses, conducted by
the LIGO-Virgo-KAGRA (LVK)
collaboration~\cite{LIGOScientific:2016aoc,LIGOScientific:2017vwq,LIGOScientific:2018mvr, LIGOScientific:2020ibl},
binary pulsars~\cite{Damour:1996ke,Yunes:2013dva,Seymour:2018bce}, and other
strong-field tests typically impose constraints on the length scale of new
couplings at the order of kilometers which is the gravitational radius of
solar-mass black
holes~\cite{Danchev:2021tew,Wong:2022wni}.
This excludes that the tachyonic instability occurs for supermassive black
holes.

In this letter, we present the first example of a theory in which the instability, and non-uniqueness, are limited to a finite range of black-hole masses that can be chosen to be supermassive. Moreover, our model incorporates a mechanism that prevents the scalarization of neutron stars, thus satisfying stringent constraints, and that resolves the incompatibility between scalarization models and cosmology \cite{Anson:2019uto,Anson:2019ebp}. The key novelty of our work, which sets it apart from other scalarization models with bounded mass windows \cite{Doneva:2017bvd,Silva:2017uqg,Silva:2018qhn}, is that in our case scalarization can occur on supermassive black hole scales. Our findings carry significant implications, as they suggest the potential existence of scalarized black holes across a spectrum of mass scales that can, for instance, be probed with the Laser Interferometer Space Antenna (LISA)~\cite{amaro2017laser,Barausse:2020rsu}, and with very-long-baseline-interferometry observations by the Event Horizon Telescope (EHT) collaboration~\cite{EventHorizonTelescope:2019ggy,EventHorizonTelescope:2020qrl,EventHorizonTelescope:2021dqv,EventHorizonTelescope:2022xqj,Ayzenberg:2023hfw}.

\noindent{\bf{\em (In)stability of the GR black hole.}}
Following typical scalarization scenarios \cite{Doneva:2022ewd} we start by considering a (yet unspecified) theory with at least one scalar field $\phi$ and with a GR vacuum, i.e., solutions obeying Einstein's equations are solutions of the new theory with a trivial (constant) profile for the scalar field $\phi=\bar{\phi}$. Additionally, scalar perturbations around the GR vacuum, $\phi=\bar{\phi} + \delta \phi$, are assumed to obey a generalized Klein-Gordon equation
\begin{equation}
    \left(\Box - \mu_{\rm eff}^2\right) \delta \phi = 0,
    \label{eq:perturbation}
\end{equation}
where $\mu_{\rm eff}$ plays the role of an effective mass for the scalar perturbations, which is allowed to be position dependent. It is manifest that if the effective mass squared is sufficiently negative, the system is subject to a tachyonic instability that may be quenched by non-linearities of the theory not captured by the (linearized) perturbation equation, in a phenomenon akin to the Higgs mechanism. In conventional scalarization scenarios \cite{Doneva:2017bvd,Silva:2017uqg}, and ignoring the scalar field's bare mass, the effective mass is typically taken to be $\mu_{\rm eff}^2 = - \alpha_1 \GB$, where $\alpha_1$ is a coupling constant with dimensions of length squared (in units $G=c=1$), and $\mathcal{G} = R^2 - 4R_{\mu\nu}R^{\mu\nu} + R_{\mu\nu\alpha\beta}R^{\mu\nu\alpha\beta}$.

Around a general static and spherically symmetric background metric
\begin{equation}
    \dd s^2 =
    -a(r) \, \dd t^2
    + b(r)^{-1} \, {\dd r^2}
    + r^2 ( \dd\theta^2 + \sin^2 \theta \, \dd\varphi^2 ),
    \label{eq:SSMetric}
\end{equation}
which may describe either a black hole or the spacetime of a stellar object, the scalar-field perturbations can be separated as
$\delta \phi = u(r) \exp(- \ii \omega t) Y_{\ell m}(\theta, \varphi) / r$,
where
$Y_{\ell m}$
are the spherical harmonics.
In terms of a radial coordinate $\dd r_{\ast} = \dd r/\sqrt{a b}$, the perturbation Eq.~\eqref{eq:perturbation} takes a Schr\"odinger-like form
\begin{equation}
    	\frac{\dd^2u}{\dd r_{\ast}^{2}} + \left( \omega^2 - V_{\rm eff} \right) u = 0,
	\label{eq:perturbation2}
\end{equation}
where the effective potential is
\begin{equation}
	V_{\rm eff}(r) = a \left[\frac{\ell \left(\ell +1\right)}{r^2} + \frac{1}{2r a}\frac{\dd (ab)}{\dd r} + \mu_{\rm eff}^2 \right].
\label{eq:veff}
\end{equation}
Focusing on monopolar perturbations ($\ell=0$) around a Schwarzschild black hole, for which $a = b =1-2M/r$, Refs. \cite{Doneva:2017bvd,Silva:2017uqg} showed that for $\mu_{\rm eff}^2 = - \alpha_1 \GB$, all Schwarzschild black holes with a mass $M \lesssim 1.174 \sqrt{\alpha_1}$ are tachyonically unstable. Since observations of solar mass compact objects impose constraints on the length scale $\sqrt{\alpha_1}$ at the order of kilometers \cite{Danchev:2021tew,Wong:2022wni}, this tachyonic instability is ruled out for supermassive black holes. However, as we show next, there exist models where scalarization occurs in a finite mass window, bounded from above and below, which can be taken to be in the supermassive range. These models do not conflict with current observational data because all compact objects with masses below the supermassive range ($M \lesssim 10^5 \mathrm{M}_\odot$) are expected to be described by GR solutions.

Consider the following modification to the effective mass of scalar perturbations
\begin{equation}
    \mu_{\rm eff}^2 = - \alpha_1 \GB + \alpha_2^3 \GB^2,
    \label{eq:effmass}
\end{equation}
where $\alpha_2$ is a new coupling with the same dimensions as $\alpha_1$. The salient new feature of our proposal is already visible here: the effective mass is determined by a competition between two different terms. At sufficiently low curvature, the first term dominates and the theory behaves similarly to standard scalar-Gauss-Bonnet gravity~\cite{Doneva:2017bvd, Silva:2017uqg}. At sufficiently high curvature, however, the second term dominates and, since it is always positive, prevents scalarization.

For a black hole, a sufficient condition for the existence of an unstable mode is
$\int_{-\infty}^{+\infty} V_{\rm eff}(r_{\ast})~\dd r_{\ast} < 0$~\cite{1995AmJPh..63..256B}, where $V_{\rm eff}$ is presented in Eq.~\eqref{eq:veff}. Heuristically, this condition, which would signal a bound state in quantum mechanics, is indicative of the existence of a mode with positive imaginary frequency, i.e., an exponentially growing perturbation. Computing the integral, assuming Eq.~\eqref{eq:effmass}, this condition is equivalent to
\begin{equation}
	1 - \frac{6}{5} \frac{\alpha_1}{M^2} + \frac{9}{22} \left(\frac{\alpha_2}{M^2}\right)^3 < 0,
\label{eq:instabilitycondition}
\end{equation}
where we used that $\GB = 48 M^2/r^6$ in the Schwarzschild spacetime. When $\alpha_2 = 0$, we recover the usual Gauss-Bonnet scalarization result~\cite{Doneva:2017bvd, Silva:2017uqg}. Instead, when $0<\alpha_2/\alpha_1 < \frac{4\times 11^{1/3}}{5 \times 3^{2/3}} \approx 0.855$, the inequality~\eqref{eq:instabilitycondition} is respected between two positive real values of $M$. There is thus a finite mass window, bounded from above and from below, within which the Schwarzschild solution must be unstable.

The inequality~\eqref{eq:instabilitycondition} is only a sufficient condition for instability. To conclusively establish instability, we use the S-deformation method~\cite{Kimura:2018eiv,Kimura:2018whv}. For perturbations as in Eq.~\eqref{eq:perturbation2}, this method establishes the linear stability of a black hole and amounts to showing that a smooth deformation function $S(r)$ exists such that
\begin{equation}
	\frac{\dd S}{\dd r} = \frac{S(r)^2-V_{\rm eff}(r)}{1 - {2M}/{r}}.
	\label{eq:Sdef}
\end{equation}
For each choice of $\alpha_1$, $\alpha_2$ and $M$, we solve Eq.~\eqref{eq:Sdef} numerically, imposing the boundary condition $S(2M)=0$.
Our results are presented in Fig.~\ref{fig:mplot}, where the hatched region shows
the Schwarzschild instability region, where we could not find a deformation function. For each value $\alpha_2/\alpha_1>0$, this corresponds to a mass window bounded from above and below. We have also numerically solved the perturbation Eq.~\eqref{eq:perturbation2} in the Schwarzschild background, for a time-independent scalar (i.e., for~$\omega=0$). We expected to find solutions if the scalarized branch connects continuously to the Schwarzschild branch;
we were not able to find solutions to the perturbation equation for masses near the lower bound of the mass range.
Further, we found no solutions to the perturbation equation for values $\alpha_2/\alpha_1 \gtrsim 0.902$, indicating that scalarized black holes cease to exist past this value for the ratio of the couplings.

\noindent{\bf{\em A complete model.}}
So far, all our analysis has relied on a perturbative treatment around a fixed GR background
In addition, we now provide an exemplary,
complete model that has both GR solutions with a trivial scalar field, and an equation for scalar perturbations that follows Eq.~\eqref{eq:perturbation} with an effective mass given by Eq.~\eqref{eq:effmass}.
To this end, we consider the following action
\begin{equation}
    \begin{aligned}
        S = \frac{1}{16\pi}
        \int \dV \,
        \bigg[ R - (\partial \phi)^2 + F\left(\phi\right) f(\GB) \bigg],
        \label{eq:actiongravity}
    \end{aligned}
\end{equation}
where $F$ and $f$ are free functions of the scalar-field and Gauss-Bonnet invariant respectively. In what follows, we take $f(\GB) = - \alpha_1 \GB + \alpha_2^3 \GB^2$, similarly to the effective mass in Eq.~\eqref{eq:effmass}. While the equations of motion for the action in Eq.~\eqref{eq:actiongravity} are fourth-order in derivatives, the model is not pathological. Using well-known techniques~\cite{Rodrigues:2011zi}, the $f(\GB)$ part of the action can be rewritten as
\begin{equation}
\begin{aligned}
	S = \frac{1}{16\pi}
    \int \dV \,
    \bigg[ &R - (\partial \phi)^2 + \alpha_1 F\left(\phi\right) \GB\\& - 2 \alpha_2^3 F\left(\phi\right)
    \left( \psi\GB - \frac{\psi^2}{2} \right) \bigg],
	\label{eq:actionST}
\end{aligned}
\end{equation}
where $\psi$ is real scalar field with the same dimension as the Gauss-Bonnet invariant, inverse length to the fourth. The equations of motion resulting from the action~\eqref{eq:actionST} are
\begin{equation}
\begin{aligned}
    	G_{\mu \nu} &= \partial_\mu \phi \partial_\nu \phi - \frac{1}{2}g_{\mu \nu}\left[ (\partial \phi)^2 - \alpha_2^3 \psi^2 F\left(\phi\right) \right] \\
        &\quad - 4 \, {}^*R^*_{\mu \alpha \nu \beta} \nabla^\alpha \nabla^\beta \left[ \left( -\alpha_1 + 2 \alpha_2^3 \psi \right) F\left(\phi\right) \right],
     \label{eq:EOMmetric}
\end{aligned}
\end{equation}
\begin{equation}
	\Box \phi = \left[ -\alpha_1\GB + 2 \alpha_2^3 \left( \psi \GB - \frac{\psi^2}{2} \right) \right] \frac{F'\left(\phi\right)}{2},
 \label{eq:EOMphi}
\end{equation}
\begin{equation}
    \psi - \GB = 0,
    \label{eq:EOMpsi}
\end{equation}
where ${}^*R^*_{\mu \alpha \nu \beta}$ is the double dual of the Riemann tensor. The equations \eqref{eq:EOMmetric}, \eqref{eq:EOMphi} and \eqref{eq:EOMpsi} are manifestly second order in all fields, the metric, $\phi$ and $\psi$, and thus our model belongs to a bi-scalar extension~\cite{Horndeski:2024hee} of Horndeski's theory~\cite{Horndeski:1974wa}, and is ghost-free.
The scalar $\psi$ is such that the model in Eq.~\eqref{eq:actionST} is on-shell equivalent to the one in Eq.~\eqref{eq:actiongravity}.

We require that the coupling function $F\left(\phi\right)$ satisfies
\begin{equation}
    F(0)=0, \qquad F'\left(0\right) = 0, \qquad F''\left(0\right)>0.
    \label{eq:conditions}
\end{equation}
The first two conditions impose that vacuum solutions of GR are solutions of the theory, when the scalar field takes the constant value $\phi=\bar{\phi}=0$. The third condition results in a tachyonic instability that we discuss in detail below. Without loss of generality, we assume $F''\left(0\right) = 2$. Other values of $F''\left(0\right)$ can be absorbed into a redefinition of $\alpha_1$ and $\alpha_2$. For small perturbations $\delta\phi$ around $\phi=0$, linearizing Eq.~\eqref{eq:EOMphi} leads to the perturbation Eq.~\eqref{eq:perturbation} with effective mass given by \eqref{eq:effmass}. We stress that the 
above model is presented as a simple working example
of a model with a scalar-perturbation equation equivalent to Eq.~\eqref{eq:perturbation} with an effective mass as in Eq.~\eqref{eq:effmass}.
In contrast, the presented mechanism is neither unique nor restricted to scalar-tensor theories.

\noindent{\bf{\em Scalarized black holes.}}
We will now establish the existence of a new branch of scalarized solutions in the mass window where the Schwarzschild solution is unstable for the theory \eqref{eq:actionST}.
For this, we work with a general ansatz for a spherically symmetric, static metric. To construct the (numerical) scalarized solutions, we follow the approach of Ref.~\cite{Fernandes:2022gde}, using a publicly available code developed by one of us. The numerical method, boundary conditions and extensive validation of the code are discussed in the Supplemental Material.
The Arnowitt-Deser-Misner (ADM) mass $M$ follows from the asymptotic behavior of the metric function $g_{tt} \sim -1 + 2M/r + \mathcal{O}\left(r^{-2}\right)$. The scalar $\phi$ with scalar charge $Q$ behaves as $\phi \sim Q/r + \mathcal{O}\left(r^{-2}\right)$ near spatial infinity.
The entropy of the black hole does not follow an area law, but can be defined as an integral over the horizon~\cite{Wald:1993nt,Iyer:1994ys}
\begin{equation}
    S = \frac{A_H}{4} + \frac{1}{4} \int_\mathcal{H} \dd^2 x \, \sqrt{\gamma} \, \left(\alpha_1 - 2\alpha_2^3 \psi \right) F(\phi) \Tilde{R},
    \label{eq:entropy}
\end{equation}
where $\gamma$ is the determinant of the induced metric on the horizon~$\mathcal{H}$, $\Tilde{R}$ is its Ricci scalar, and $A_H$ is the area of the event horizon.
To construct the numerical solutions, we chose the quadratic-exponential coupling commonly used in the literature~\cite{Doneva:2017bvd}, which satisfies the conditions~\eqref{eq:conditions} and is known to produce stable solutions in the scalar-Gauss-Bonnet model
\begin{equation}
\label{eq:coupling_F}
    F(\phi) = \frac{1}{6} \left(1 - e^{-6 \phi^2} \right).
\end{equation}

\begin{figure}[]
	\centering
    \includegraphics[width=0.88\columnwidth]{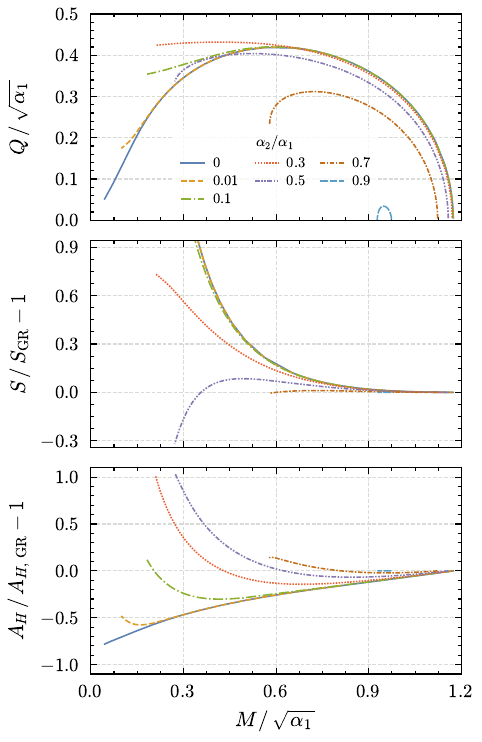}
        \caption{Scalar charge (upper panel), entropy (middle panel), and horizon area (lower panel) of the scalarized solutions as functions of $M/\sqrt{\alpha_1}$. The scalar charge is normalized to the value of the coupling, while the entropy and horizon area are compared with those of a Schwarzschild black hole with the same mass.}
	\label{fig:numerical}
\end{figure}
We have confirmed that for each nonzero value of $\alpha_2/\alpha_1$, scalarized solutions exist only within a mass window bounded from above and below, as shown in Fig.~\ref{fig:mplot}, where the red shaded region denotes the domain of existence of scalarized black holes for each coupling ratio $\alpha_2/\alpha_1$. While the bifurcation points from the Schwarzschild solution were notably in agreement with the upper bound S-deformation predictions, the termination point of the scalarized black hole branch differed, in general.
In particular, for couplings obeying $\alpha_2/\alpha_1 \lesssim 0.3$, the endpoint of the scalarized branches occurs for values of the mass higher than those obtained with the S-deformation method. This suggests a mass range where stable, spherically symmetric and
static black hole solutions are entirely absent. On the other hand, for higher $\alpha_2/\alpha_1$ the scalarized branch ends for masses smaller than those obtained with the S-deformation method, leading to a mass range where stable Schwarzschild black holes and scalarized black holes co-exist. As we approach the maximum value $\alpha_2/\alpha_1 \approx 0.902$, the endpoint of the scalarized branches also approach the lower-bound values obtained with the S-deformation method.
In contrast to typical scenarios in standard scalar-Gauss-Bonnet gravity, where branches terminate in singular solutions~\cite{Fernandes:2022kvg}, we detect no singular behavior in the solutions when inspecting the Ricci and Gauss-Bonnet scalars at the black hole horizon. Instead, we observe a decline in the accuracy of the scalarized solutions as we approach the end of the scalarized branch, and at a certain point, the code ceases to converge to a scalarized solution. Past this point, our numerical method only finds
Schwarzschild black holes. We believe this is a physical (rather than numerical) feature given the extensive testing of the code -- see the Supplemental Material.

As shown in Fig.~\ref{fig:numerical}, the entropy of the scalarized solution is greater than that of a corresponding Schwarzschild black hole, except in proximity to the termination of the existence domain in certain instances, such as for $\alpha_2/\alpha_1 = 0.5$ and $\alpha_2/\alpha_1 = 0.7$. This trend signifies an entropic preference for scalarized solutions in most cases. Notably, the physical quantities exhibit non-monotonic deviations from those of GR, with regions where they are both smaller and larger. This is evident, for instance, in the variation of the horizon area as shown in Fig.~\ref{fig:numerical}. The scalar charge $Q$ normalized to the length scale $\sqrt{\alpha_1}$ is also shown in Fig.~\ref{fig:numerical}. Plots for other physical quantities of interest in astrophysical contexts are shown in the Supplemental Material.
We have also studied rotating black hole solutions of the theory \eqref{eq:actionST}, following the approach in Ref. \cite{Fernandes:2022gde}. Our analysis shows that, qualitatively, these rotating solutions closely resemble the static ones, with the main distinction being a narrower domain of existence for the spinning configurations, see Supplemental Material for further details.

\begin{figure}[]
    \includegraphics[width=\linewidth]{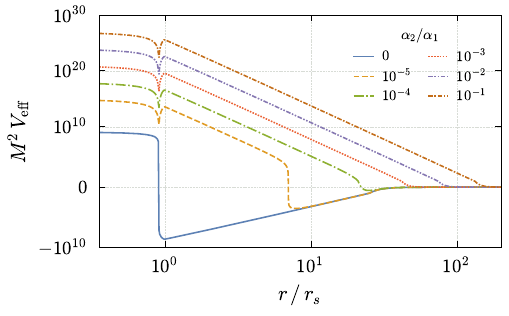}
    \caption{The dimensionless effective potential $M^2 V_{\rm eff}$ for scalar perturbations around a neutron star with $M \approx 1.4~{\rm M}_\odot$ and radius $r_s \approx 11 \, \mathrm{km}$, described by the AP4 equation of state~\cite{Akmal:1998cf}, considering various values of $\alpha_2/\alpha_1$ and $\sqrt{\alpha_1} \approx 10^6~{\rm M}_\odot$.}
	\label{fig:NSVeff}
\end{figure}

\noindent{\bf{\em Scalarization in neutron stars and other less compact
objects.}}
We do not, in general, expect scalarization to occur in neutron stars.
We have solved the Tolman-Oppenheimer-Volkoff equations for a neutron star model with $M=1.4~\textrm{M}_\odot$, employing the AP4 equation of state~\cite{Akmal:1998cf}. In Fig.~\ref{fig:NSVeff}, we plot the effective potential \eqref{eq:veff} for scalar perturbations around this neutron star solution, considering various values of $\alpha_2/\alpha_1$ and assuming $\sqrt{\alpha_1} \approx 10^6~\textrm{M}_\odot$.
With increasing $\alpha_2/\alpha_1$, the negative regions in the effective potential become progressively weaker and shift further away from the surface of the neutron star. The effective potential within the star is overwhelmingly positive if $\alpha_2 \neq 0$. Consequently,
we expect that scalarization of neutron stars can be avoided, at least if both couplings are sufficiently large, i.e., $\alpha_1\gg~\textrm{M}_\odot$ and $\alpha_2\gg~\textrm{M}_\odot$. This trend remains true for the other equations of state we have explored, such as SLy \cite{Douchin:2001sv}.

Defining the compactness of a body with ADM mass $M$ and radius $r_s$ as $C = 2M/r_s$, the effective potential in Eq.~\eqref{eq:veff} is proportional to an overall factor of $C^3$ in the exterior region ($r\geq r_s$) where the metric is Schwarzschild,
\begin{equation*}
    V_{\rm eff} = \frac{C^3}{M^2}\frac{\left(r/r_s-C\right)}{ 4(r/r_s)^4}\left[ 1- 2\sqrt{3}\alpha_1 \sqrt{\GB} + 2\sqrt{3} \left( \alpha_2 \sqrt{\GB}\right)^3 \right].
\end{equation*}
We observe that when two objects possess different masses, but their Gauss-Bonnet curvature near the surface is of the same order of magnitude, the effective potential of the less compact object is suppressed by a factor of~$C^3$. This suppression makes it more challenging for the less compact object to undergo scalarization. Therefore, astrophysical bodies like the Sun (for which $C \sim 10^{-6}$), exhibit an extremely weak effective potential near their surface, even if black holes (for which $C =1$) with the same surface Gauss-Bonnet curvature are scalarized.
Consequently, we do not anticipate scalarization to occur, for instance, within the solar system. Similarly, we do not expect that laboratory experiments which are sensitive to curvatures of the same order as the horizon curvature of a supermassive black hole, can probe the scalarized regime.

\noindent{\bf{\em Stability of cosmological evolution.}}
Deviations from GR are constrained by cosmological observations. We therefore check that the Friedmann-Robertson-Walker solution is unaffected by the same tachyonic instability that scalarizes black holes.
In a Friedmann-Robertson-Walker spacetime, the Gauss-Bonnet invariant is  
$\mathcal{G} = 24 H^2 \ddot{a}/a$, where $a$ is the scale factor, $H=\dot{a}/a$ is the Hubble parameter, and dots indicate time derivatives. In a decelerating universe ($\ddot{a} < 0$), such as during matter or radiation domination eras, the Gauss-Bonnet invariant is negative, making the squared effective mass of the scalar perturbations positive; see Eq.~\eqref{eq:effmass}. However, in an accelerating universe, such as during inflation or during dark energy domination, the squared effective mass may turn negative.
For a constant Hubble rate, $\ddot{a}/a= H^2$ and thus $\mu_{\rm eff}^2 = 24 H^4 \left(- \alpha_1+ 24 \alpha_2^3 H^4 \right)$. Whether the first or the second term dominates depends on the choice of $\alpha_{1,2}$ and the value of $H$. To be of astrophysical relevance, our model should have values of $\sqrt{\alpha_1}$ between the gravitational radii of stellar-mass and supermassive black holes. This implies that $\alpha_1$ ranges between $\mathcal{O}\left(10^{36} {\rm GeV^{-2}}\right)$ and $\mathcal{O}\left(10^{48} {\rm GeV}^{-2}\right)$, and $\alpha_2 \sim \alpha_1$.

In the late universe, where $H = H_0 \sim 10^{-42} \, \mathrm{GeV}$,
the mass becomes independent of $\alpha_2$ and recovers a well-known tachyonic instability of standard scalarization models~\cite{Anson:2019uto}. However, the corresponding timescale $t_{\rm inst}$ compared to the age of the universe $t_0$ is 
$t_{\rm inst}/t_0 \sim H_0 / \mu_{\rm eff} \sim 1 / (\alpha_1 H_0) \gtrsim  10^{38}$; 
thus this instability is in practice irrelevant~\cite{Anson:2019uto}.

In the early universe, assuming the energy scale of inflation to be approximately $H \sim 10^{13} \, \mathrm{GeV}$,
the $\GB^2$ term dominates. As a result, the effective mass squared remains positive throughout inflation.
This is a major difference to standard scalarization models, in which the inflationary phase is tachyonically unstable with a short time scale \cite{Anson:2019uto, Anson:2019ebp}. Our model is the first model of spontaneous scalarization that is not subject to this catastrophic instability.

\noindent{\bf{\em Conclusions and outlook.}}
There is a widely accepted expectation that, if the Kerr hypothesis breaks down, it does so above a critical curvature threshold. We present the first paradigmatic example of a theory in which this expectation does not hold true, because black-hole uniqueness is broken -- and the Schwarzschild solution is unstable --  in a \emph{finite} and bounded window of horizon curvature, or, equivalently, black-hole mass. Moreover, our theory cures the catastrophic instability in the early universe that plagues standard scalarization models.
The scalarization mass window depends on the two coupling constants $\alpha_1$ and $\alpha_2$ in the action.
If both coupling constants are of the same order of magnitude and there is no hierarchy of length scales, the window is narrow and targets a specific black-hole mass scale $M \sim \sqrt{\alpha_1} \sim \sqrt{\alpha_2}$. Conversely, it is necessary to introduce a hierarchy of scales into the theory in order to obtain an expansive window.

In GR, the presence of new ultralight bosonic matter fields can trigger a superradiant instability on Kerr black holes and lead to supermassive hairy solutions~\cite{Herdeiro:2014goa,Herdeiro:2016tmi}. However, our model and its scalarized solutions are distinct from this scenario. In our framework, black holes do not require spin to develop hair, and deviations from the Kerr metric indicate new physics beyond GR rather than the existence of new matter fields. Moreover, the nature of the instability causing deviations from the Kerr black hole is different between the two models.

Dependent on the value of $\alpha_1\sim\alpha_2$, different observational channels can constrain the theory.
First, if the mass window lies within the solar-mass range, our theory can be constrained by LVK observations~\cite{LIGOScientific:2016aoc,LIGOScientific:2017vwq,LIGOScientific:2018mvr, LIGOScientific:2020ibl}.
For instance, an equal-mass binary may start out as a binary of two GR black holes, but end up in the mass window in which the merger remnant may be scalarized. This is in contrast to other scalarization scenarios in the literature and could lead to a detectable mismatch between the expected and the observed post-merger phase of the gravitational-wave signal.

Second, if $\alpha_2$ and $\alpha_1$ are such that solar-mass black holes are described by the Schwarzschild solution, but supermassive black holes are not, LVK observations are insensitive to the modifications to GR and observations of supermassive black holes constitute the only pathway to constrain the theory.
We expect that our model can already be constrained with
existing EHT observations~\cite{EventHorizonTelescope:2019ggy,EventHorizonTelescope:2020qrl,EventHorizonTelescope:2021dqv,EventHorizonTelescope:2022xqj} and we will address this in future work. Given the factor of roughly $10^3$ between the mass of Sgr~A* and M87*, it may well be that only one of them is scalarized, because a hierarchy between $\alpha_1$ and $\alpha_2$ must be introduced in order to scalarize both.

Future upgrades of the EHT \cite{Ayzenberg:2023hfw, Johnson:2023ynn} and planned space-based gravitational-wave observatories like LISA \cite{amaro2017laser,Barausse:2020rsu}, TianQin \cite{TianQin:2015yph} or Taiji \cite{Hu:2017mde} may also probe the theory.
As a first step in this direction, one may focus on extreme-mass-ratio-inspirals (EMRIs), in which a small (nonscalarized) black hole orbits a large (scalarized) black hole because EMRIs are an important observational target for LISA.
Another possibility, is to analyze the coalescence of massive black-hole binaries, which may be detected by LISA with signal-to-noise ratio
reaching thousands~\cite{Klein:2015hvg}.
To determine whether stellar orbits and gas dynamics near supermassive black holes as well as x-ray observations also provide constraints, one should characterize deviations in a post-Newtonian expansion.
If supermassive black holes surpass the instability threshold due to accretion, the resulting spontaneous transition may also result in observable signatures.
In addition, scalarization may affect the stochastic gravitational-wave background from a population of supermassive black holes, leaving an imprint on pulsar-timing-array data~\cite{NANOGrav:2023gor}.

As $\alpha_2/\alpha_1$ decreases, the scalarization window widens (see Fig.~\ref{fig:mplot}) and the theory may be constrained at various mass scales.
Moreover, we find indications for a mass gap, i.e., a range of black-hole masses for which no stable branch of spherically symmetric and stationary solutions exists. To further understand the different black-hole branches and whether the mass gap persists beyond spherical symmetry, we plan to study rotating black-hole solutions. At nonzero spin, minimally and nonminimally coupled scalar fields can trigger new instabilities due to superradiance~\cite{Herdeiro:2014goa,Herdeiro:2016tmi} and spin-induced scalarization~\cite{Dima:2020yac,Herdeiro:2020wei,Berti:2020kgk}. Preliminary work in this direction is ongoing, with some initial results available in the Supplemental Material. In general, the qualitative behavior of these rotating solutions closely mirrors that of the static ones. A full investigation of rotating black holes of the theory will be addressed in a future publication by us.

Finally, the upper curvature bound may even be taken to cosmological scales. The present theory might provide a screening mechanism which hides cosmological modifications from solar-system (and even galactic) observations. The latter application assumes that our results on black-hole (non)uniqueness carry over to cosmological spacetimes.

\noindent{\bf{\em Acknowledgements.}}
%
We thank Thomas Sotiriou and the anonymous referees for valuable comments on the manuscript.
AE and PGSF are supported by a grant (29405) from VILLUM fonden.
HOS acknowledges funding from the Deutsche Forschungsgemeinschaft
(DFG)~-~Project No.:~386119226.
This work is funded by the Deutsche Forschungsgemeinschaft (DFG, German Research Foundation) under Germany’s Excellence Strategy EXC 2181/1 - 390900948 (the Heidelberg STRUCTURES Excellence Cluster).

\bibliography{biblio}

\onecolumngrid
\newpage
\setcounter{equation}{0}
\renewcommand{\theequation}{\roman{equation}}
\section*{Supplemental Material}
\subsection*{Numerical method}
We work with a general ansatz for a spherically symmetric, static metric, in isotropic coordinates,
\begin{align}
    \dd s^2 = -f \mathcal{N}^{2} \, \dd t^2 + \frac{g}{f} \left(\dd \rho^2 + \rho^2 \, \dd\theta^2 + \rho^2 \sin^2\theta \, \dd\varphi^2\right),
    \label{eq:metric}
\end{align}
where $\mathcal{N} = 1-\rho_H/\rho$ and $f$ and $g$ are functions of $\rho$, with $\rho_H$ the coordinate location of the event horizon.
In our numerical setup, we utilize the compactified radial coordinate $x=1-2\rho_H/\rho$, which maps the interval $[\rho_H, \infty[$ to $[-1, 1]$.
The boundary conditions we impose are as follows:
at the horizon ($x=-1$), our functions obey $f-2 \partial_x f = g+2 \partial_x g = \partial_x \phi = \partial_x \psi = 0$. To ensure asymptotic flatness, we impose that $f=g=h=1$, and $\phi = \psi = 0$ at $x=1$.
To solve the differential-algebraic system resulting from the field
equations~(2),~(3) and~(4) in the main text,
our code employs a pseudospectral method together with the Newton-Raphson root-finding algorithm (see Refs.~\cite{Dias:2015nua,Fernandes:2022gde}). We expand each of the functions in a spectral series with resolution $N_x$ in the radial coordinate $x$. The spectral series we use for each of the functions $\mathcal{F}^{(k)}=\{f,g,\phi,\psi\}$ is given by
\begin{equation}
  \mathcal{F}^{(k)} = \sum_{i=0}^{N_x-1} c_{i}^{(k)} T_i(x),
\label{eq:spectralexpansion1}
\end{equation}
where $T_i(x)$ denotes the $i$-th Chebyshev polynomial, and $c_i^{(k)}$ are the spectral coefficients of the $k$-th function.
We have typically used a resolution $N_x \sim 40$.

\begin{figure}[h!]
\includegraphics[width=0.5\linewidth]{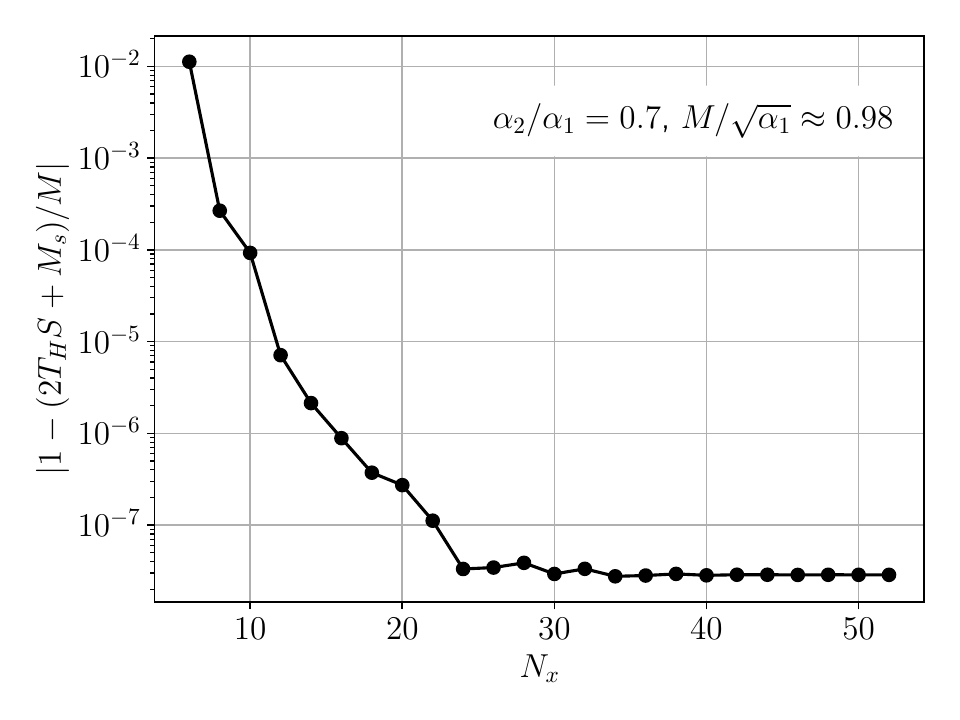}
\caption{Convergence test for a particular solution of our model with $\alpha_2/\alpha_1=0.7$ and $M/\sqrt{\alpha_1} \approx 0.98$. We observe that as the resolution $N_x$ is increased, the error in the numerical solution as estimated by the Smarr-type relation decreases exponentially until a plateau is reached where the error is estimated to be less than $\mathcal{O}\left(10^{-8}\right)$.}
\label{fig:convergence}
\end{figure}

From the physical quantities of the solutions, we can estimate the accuracy of the numerical solutions using a Smarr-type relation \cite{PhysRevLett.30.71} they should obey
\begin{equation}
    M = 2 T_H S + M_s, \qquad {\rm where } \quad  M_s = \frac{1}{4\pi} \int \dd^3x \sqrt{-g}~F(\phi)\left[\frac{\Box \phi}{F'(\phi)} + \alpha_2^3 \psi^2\right]
    \label{eq:smarr}
\end{equation}
and where $T_H = \frac{1}{2\pi \rho_H} \frac{f}{\sqrt{g}} \rvert_\mathcal{H}$,
is the Hawking temperature of the solutions. The area of the event horizon is given by $A_H = 4\pi \rho_H^2 g/f \rvert_\mathcal{H}$. This Smarr-type relation can be derived using the formalism of Refs. \cite{Iyer:1994ys,Liberati:2015xcp}, as explained in Ref. \cite{Fernandes:2022gde}.
We assessed the numerical accuracy of our solutions through multiple approaches. Firstly, we considered Eq.~\eqref{eq:smarr}, recognizing that it is susceptible to numerical errors due to numerical integration, and thus we expect the true error to be less than estimated by the Smarr relation. Additionally, we examined the order of magnitude of the last retained spectral coefficient, and the minimization of the residuals. In general, the Smarr relation yielded errors on the order of $\mathcal{O}\left(10^{-8}\right)$. However, as we approached the end of the existence domain for scalarized solutions with sufficiently large couplings, errors increased to around $\mathcal{O}\left(10^{-4}\right)$. This error pattern aligns with what is observed with our code in the standard scalar-Gauss--Bonnet case, where it underwent thorough testing~\cite{Fernandes:2022gde, Fernandes:2022kvg} and faithfully reproduced results from the existing literature~\cite{Doneva:2017bvd, Antoniou:2017acq, Silva:2017uqg, Cunha:2019dwb}. The last retained spectral coefficient typically yielded an error estimate approximately a few orders of magnitude smaller than that of the Smarr relation, while the norm of the residuals consistently dropped below $\mathcal{O}\left(10^{-10}\right)$. To demonstrate the exponential convergence of our method, in Fig. \ref{fig:convergence} we plot the error as estimated by the Smarr-type relation as a function of the resolution $N_x$ for a particular solution with $\alpha_2/\alpha_1=0.7$ and $M/\sqrt{\alpha_1} \approx 0.98$. A similar behaviour was observed for other solutions.
Finally, we observed that the bifurcation points from Schwarzschild, for each value of $\alpha_2/\alpha_1$,
agree with high-accuracy with the upper bound values derived from the S-deformation method, providing a final test to the code.

\subsection*{Physical quantities of interest of the scalarized solutions}

\begin{figure}[]
\includegraphics[width=\linewidth]{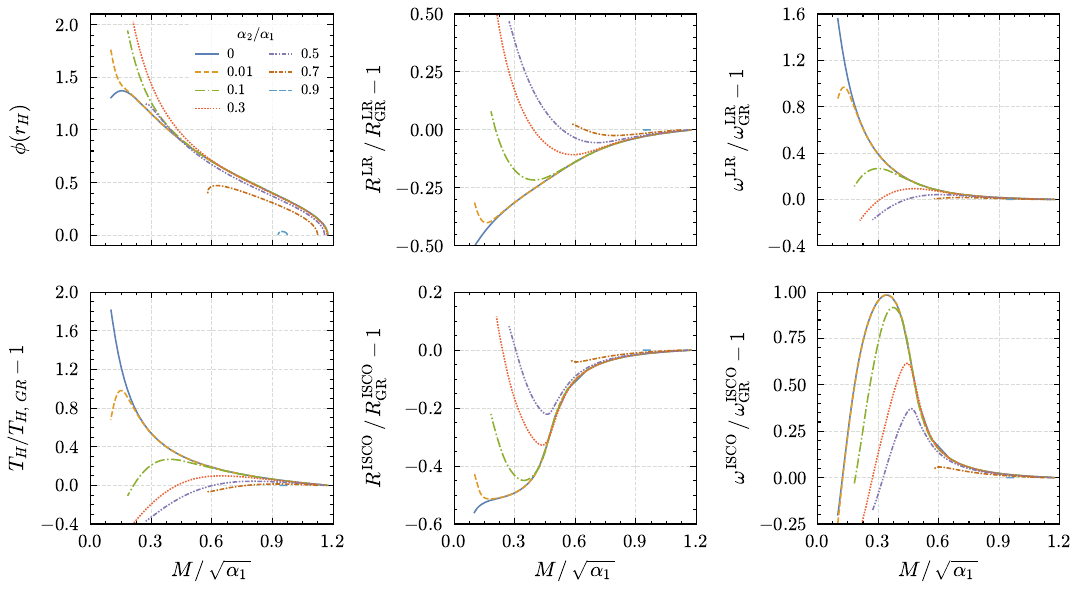}
\caption{Physical quantities characterizing scalarized solutions for a selection of $\alpha_2/\alpha_1$ values. Progressing from the columns from left to right, these include: the value of $\phi$ at the horizon and the Hawking temperature, the perimetral location of the light ring, perimetral location of the ISCO, and the geodesic frequency at the light ring, and the geodesic frequency at the ISCO. With the exception of the value of $\phi$ at the horizon, we compare all quantities to those of an equivalent Schwarzschild black hole.}
\label{fig:quantities}
\end{figure}

In Fig. \ref{fig:quantities} we plot physical quantities of interest of scalarized solutions for a sample of ratios $\alpha_2/\alpha_1$, including the value of $\phi$ at the horizon, the Hawking temperature, the perimetral location of the light ring, the perimetral location of the innermost stable circular orbit (ISCO), and the geodesic frequency at both the light ring and the ISCO. Apart from the value of $\phi$ at the horizon, we compare all these quantities to their counterparts in an equivalent Schwarzschild black hole. We observe a non-monotonic behavior, in general.

\subsection*{Rotating black holes}
\begin{figure}[]
\includegraphics[width=0.5\linewidth]{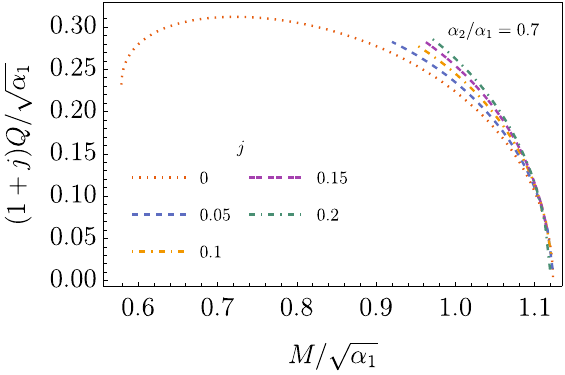}
\caption{Scalar charge of rotating scalarized black holes with dimensionless spins $j$ up to $0.2$ for the case $\alpha_2/\alpha_1=0.7$. The scalar charge is scaled by a factor of $(1+j)$ in order to discern between the different curves.}
\label{fig:rotating}
\end{figure}

Because of the relevance of spin in astrophysical black holes, we have started a preliminary study of rotating scalarized black holes of the theory. To obtain the rotating solutions we follow closely the numerical method of Ref. \cite{Fernandes:2022gde}. Some of these results can be found in Fig. \ref{fig:rotating} for the case $\alpha_2/\alpha_1 = 0.7$. Overall we have found that spinning scalarized black holes exist, and the curves for their physical quantities follow closely those of static black holes, with the domain of existence shrinking with increasing spin.

\end{document}